\documentclass{article}
\usepackage{dcase2021,amsmath,graphicx,url,times,booktabs, tabularx}
\usepackage{array,multirow}
\usepackage{amsfonts}
\usepackage{svg}
\usepackage{bm}
\usepackage{hyperref}

\title{Improving the Performance of Automated Audio Captioning via Integrating the Acoustic and Semantic Information}

\name{Zhongjie ye$^{1}$,
      Helin Wang$^{1}$,
      Dongchao Yang$^{1}$, 
      Yuexian Zou$^{1,2,}\sthanks{Yuexian Zou is the corresponding author.}$
      }
\address{$^1$ ADSPLAB, School of ECE, Peking University, Shenzhen, China\\          
        $^2$ Peng Cheng Laboratory, Shenzhen, China\\
        \{zhongjieye@stu.pku.edu.cn, wanghl15@pku.edu.cn \\
         dongchao98@stu.pku.edu.cn, zouyx@pku.edu.cn\} }

\begin{document}

\ninept
\maketitle

\begin{sloppy}

\begin{abstract}
Automated audio captioning (AAC) has developed rapidly in recent years, involving acoustic signal processing and natural language processing to generate human-readable sentences for audio clips. The current models are generally based on the neural encoder-decoder architecture, and their decoder mainly uses acoustic information that is extracted from the CNN-based encoder. However, they have ignored semantic information that could help the AAC model to generate meaningful descriptions. This paper proposes a novel approach for automated audio captioning based on incorporating semantic and acoustic information. Specifically, our audio captioning model consists of two sub-modules. (1) The pre-trained keyword encoder utilizes pre-trained ResNet38 to initialize its parameters, and then it is trained by extracted keywords as labels. (2) The multi-modal attention decoder adopts an LSTM-based decoder that contains semantic and acoustic attention modules. Experiments demonstrate that our proposed model achieves state-of-the-art performance on the Clotho dataset. Our code can be found 
at \url{https://github.com/WangHelin1997/DCASE2021_Task6_PKU}.

\end{abstract}


\section{Introduction}
\label{sec:intro}
Automated audio captioning (AAC) is a cross-modal task of generating a natural language description for an audio clip. It is different from audio tagging (AT), acoustic scene classification (ASC) and automatic speech recognition. The purpose of AAC is not only to analyze acoustic scenes, events, and concepts in a given audio clip, but also to find the relationships among them to produce human-readable sentences. Applications of automated audio captioning are diverse such as assisting the hearing impaired people by converting audio signals into a text, and content-based audio retrieval task which uses the free-form natural language queries to retrieve the audio \cite{oncescu2021audio}. 


AAC has aroused a lot of interest among researchers since the Detection and Classification of Acoustic Scenes and Events (DCASE) 2020 challenge. Nowadays, the mainstream framework is based on neural encoder-decoder systems which have achieved success in some relevant fields such as image captioning \cite{xu2015show}. The current AAC models consist of a convolutional neural network (CNN) encoder and a recurrent neural network (RNN) (or Transformer) decoder with an attention mechanism. The inputs used could be log-mel energies, Mel-Frequency Cepstral Coefficients(MFCCs), or other acoustic features which are extracted from raw audio clips. They are firstly encoded by a CNN encoder into a set of feature vectors. Then, they are decoded into sentences by an RNN-based or Transformer-based decoder with (or without) an attention mechanism. 

Over past few years, there are amounts of methods proposed in AAC task \cite{wu2019audio, wu2020audio, wang2020automated, xu:2021:ICASSP:02, koizumi2020transformer} based on neural encoder-decoder systems. M. Wu \textit{et al.} \cite{wu2019audio} straightly takes the mean of the feature vectors that are the outputs of the encoder in the time dimension, and uses them as the input of the decoder. H. Wang \textit{et al.} \cite{wang2020automated} proposed a temporal attention mechanism in the decoder, which could utilize more acoustic information for each time step. In contrast to previous work in AAC, Y. Wu \textit{et al.} \cite{wu2020audio} and X. Xu \textit{et al.} \cite{xu:2021:ICASSP:02} explore transfer learning method to help AAC models to get better performance. The strategy of their proposed methods could be divided into two stages. In the first stage, a tagging system is pre-trained by ASC or AT task. Then the parameters of the audio encoder are initialized by the pre-trained tagging system. In the second stage, the whole AAC model is trained end-to-end by minimizing the cross-entropy (CE) loss. With these methods mentioned above, they generally only consider acoustic information while ignoring semantic information when the AAC model generates sentences. Specifically, the semantic information could contain keywords that are from the encoder, previously predicted words in the decoding time, and so on. In this paper, we introduce semantic information with acoustic information to assist the decoder to generate higher quality sentences. Furthermore, to better make use of semantic and acoustic information, we propose a novel multi-modal attention mechanism. In summary, our contributions are as follows:


\begin{enumerate}
\item We propose a \textbf{m}ulti-modal \textbf{a}ttention-based \textbf{a}udio \textbf{c}aptioning model with a pre-trained keyword encoder, named \textbf{MAAC}. It could utilize both acoustic and semantic information to generate the description. The semantic information includes keywords from the pre-trained keyword encoder and the previously decoding information from the decoder.
\item Our MAAC achieves a new state-of-the-art performance on the Clotho dataset. We present the ablation analysis of the components of our MAAC and demonstrate that semantic information could improve the performance of the AAC model.

\end{enumerate}

\begin{figure*}[t]
  \centering
  \centerline{\includegraphics[width=\linewidth]{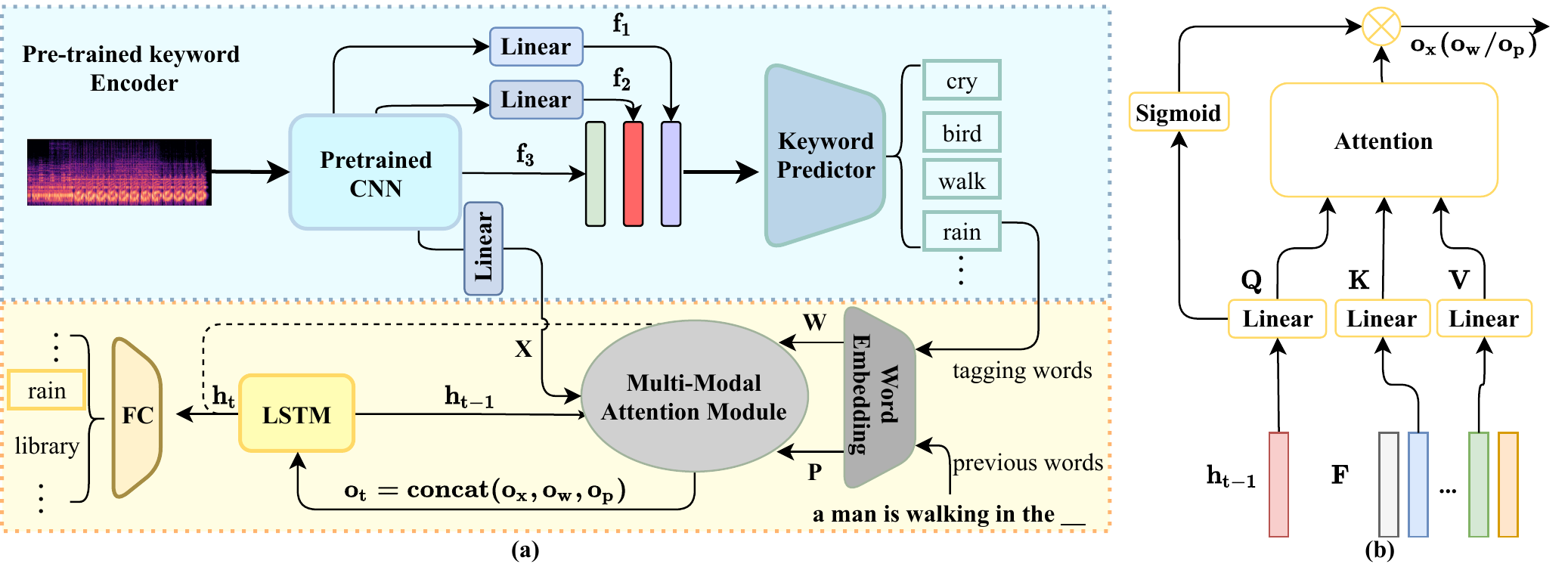}}
  \caption{(a) Our proposed MAAC includes two submodules: the pre-trained keyword encoder is on the top and the LSTM-based decoder with a multi-modal attention module is on the bottom. (b) The architecture of the attention mechanism. $F$ could represent acoustic features or semantic features.}
  \label{fig:results1}
\end{figure*}
The organization of the paper is as follows. Section 2 introduces our proposed model. We present our experimental results and evaluations in Section 3. Finally, we give concluding remarks and possible future directions in Section 4.


\section{System Architecture}
\label{sec:format}
In this section, our proposed MAAC is introduced and its architecture is shown in Figure 1. Specifically, our MAAC consists of two submodules: a pre-trained keyword encoder and an LSTM-based decoder with a multi-modal attention module. In the following subsections, we will introduce details about it.

\subsection{Pre-trained Keyword Encoder}
\label{sec:keywordencoder}
The CNN encoder, which is widely used in the AAC challenge \cite{wu2020audio, wang2020automated}, plays an important role in extracting acoustic information from raw audios. In this work, we extract keywords from captions as training labels and use the pre-trained ResNet38\footnote{\url{https://github.com/qiuqiangkong/audioset_tagging_cnn}} \cite{kong2020panns} that performs well in the AudioSet dataset \cite{gemmeke2017audio} as our backbone network.

\noindent\textbf{Constructing Audio-Keyword Training Pairs} Firstly, Natural Language Toolkit (NLTK\footnote{\url{https://github.com/nltk/nltk}}) is a powerful open-source tool applied to extract words from each caption. We choose the nouns and verbs to construct the keyword table by getting rid of some useless words such as \textit{make}, \textit{go}, \textit{others}, etc. The verbs in the keyword table are transformed into their original forms and the nouns are not changed, because plural forms of the nouns have different meanings. Then, we choose $N$ keywords with the highest frequency from the modified keyword table and use them as labels for pre-training.

We combine all the keywords from the 5 captions of each audio clip to form the training label which is a multi-hot vector. Each word of captions is transformed into its original forms according to the above rules. When a word occurs in the keyword table, the corresponding position of the multi-hot vector is set to 1, otherwise 0.

\begin{table*}[h]
\caption{Single-model performances on the Clotho \cite{drossos2020clotho} evaluation splits in the CE and RL training period. B1, B4, RG, ME, CD, SP, and SD denote BLEU-1, BLEU-4, METEOR, ROUGE-L, CIDEr-D, SPICE, and SPIDEr, respectively. For all metrics, higher values indicate better performance.}
\begin{tabular}{c|ccccccc|ccccccc}
    \toprule 
     & \multicolumn{7}{c|}{{  Cross-entropy}} & \multicolumn{7}{c}{{  CIDEr-D optimization}} \\
     \midrule 
    Model &
      \multicolumn{1}{l}{{   B1}} &
      \multicolumn{1}{l}{{   B4}} &
      \multicolumn{1}{l}{{   RG}} &
      \multicolumn{1}{l}{{   ME}} &
      \multicolumn{1}{l}{{   CD}} &
      \multicolumn{1}{l}{{   SP}} &
      \multicolumn{1}{l|}{{   SD}} 
      &
      \multicolumn{1}{l}{{   B1}} &
      \multicolumn{1}{l}{{   B4}} &
      \multicolumn{1}{l}{{   RG}} &
      \multicolumn{1}{l}{{   ME}} &
      \multicolumn{1}{l}{{   CD}} &
      \multicolumn{1}{l}{{   SP}} &
      \multicolumn{1}{l}{{   SD}} 
       \\ 
    \midrule 
    Baseline \cite{drossos2020clotho}   & 37.8 &  1.7 & 26.3 & 7.8 & 7.5 & 2.8 & 5.1   & - & - & - & - & - & - & -\\
    TAM \cite{wang2020automated}   & 48.9 & 10.7 & 32.5 & 14.8 & 25.2 & 9.1 & 17.2 & - & - & - & - & - & - & -  \\
    TM \cite{wu2020audio} & 53.4 &  15.1 & 35.6 & 16.0 & 34.6 & 10.8 & 22.7 & - & - & - & - & - & - & -  \\
    UNIS's model \cite{xinhao2021_t6} & - & - & - & - & - & - & - & 62.5 & 17.8 & 40.1 & 17.6 & 42.8 & 12.6 & 27.7   \\
    SJTU's model \cite{xu2021_t6} & 56.5 &  15.5 & 37.4 & 17.4 & 39.9 & 11.9 & 25.9 & 64.0 & 16.3 & 40.4 & 17.8 & 44.9 & 12.3 & 28.6   \\
    
    \midrule 
    MAAC (Ours)        & 57.7 & 17.4 & 37.7 & 17.4 & 41.9 & 11.9 & 26.9  & \textbf{64.8} & \textbf{18.1} & \textbf{40.8} & \textbf{19.0} & \textbf{49.1} & \textbf{13.1} & \textbf{31.1} \\
    \bottomrule
\end{tabular}
\end{table*}

\noindent\textbf{Training the Keyword Encoder} As Figure 1 illustrates, the pre-trained ResNet38 is used as our backbone, which consists of 6 convolutional blocks. We refine it with a feature hierarchy structure to combine multi-level features, \textit{i.e.} the features after the third, fourth, and last convolution block. Then all of them are passed into different linear layers after the global average pooling (GAP) method to obtain $f_{1}$, $f_{2}$ and $f_{3}$. Finally, we use them to obtain the predictions $\hat{y}\in\mathbb{R}^{N}$ and $N$ is the number of keywords.
\begin{equation}
    \hat{y} = \sigma(Linear(concat(f_1,f_2,f_3)))
\end{equation}
where $\sigma$ denotes sigmoid activation function. Given the ground-truth $y\in\mathbb{R}^{N}$, the pre-trained keyword encoder could be optimized by minimizing the binary cross-entropy loss:
\begin{equation}
    \mathcal{L}_{bce}(y,\hat{y}) = -\frac{1}{N}\sum_{i=1}^N y(i)log\ \hat{y}(i) \label{con:celoss}
\end{equation}


\begin{table}[t]
    \centering  
    \caption{Settings and results of ablation studies. The results are reported after CE training stage. SAM denotes the semantic attention module.}
    \begin{tabular}{l|ccc}
    \toprule 
    \multicolumn{1}{l|}{{   Model}} &
      \multicolumn{1}{l}{{   B4}} &
      \multicolumn{1}{l}{{   CD}} &
      \multicolumn{1}{l}{{   SD}}  \\
    \midrule 
    Base  & 16.5 & 40.6  & 26.4 \\
    \midrule 
    + Previously predicted words  & 17.1 &  41.1 &  26.4 \\
    + Keywords  & \multicolumn{3}{c}{{can not converge}} \\
    + Both (w/o sharing SAM)  & 16.8 & 41.1  & 26.7 \\
    \midrule 
    proposed MAAC  & \textbf{17.4} & \textbf{41.9}  & \textbf{26.9} \\
    \bottomrule
\end{tabular}
\end{table}

\subsection{Multi-modal Attention Decoder}
\label{sec:attention}
Unlike the existing audio captioning models, we further incorporate acoustic with semantic information into generating captions: we propose a multi-modal attention module to incorporate them. The high-level representation of acoustic information denoted as $\bm{X} = \{x_1,...,x_L\} \in \mathbb{R}^{L \times C_{1}}$, is the output of a linear layer whose input is the output of the last convolution block of the pre-trained keyword encoder. The semantic features contain the keywords $\bm{W} = \{w_1,...,w_K\}$ that is the $K$ outputs of the pre-trained keyword encoder, and the previously predicted words $\bm{P} = \{p_1,...,p_{t-1} \}$ that contain all the generated words before time step t. Both of them are transformed into continuous vectors by a randomly initialized embedding layer $\textbf{Emb}$, $\bm{W}\in\mathbb{R}^{K \times C_{2}}$ and $\bm{P}\in\mathbb{R}^{(t-1) \times C_{2}}$. The implementation process of the multi-modal attention module is as follows.

Firstly, all of them are transformed into the same latent space, where $X$ is turned to $\bm{\hat{X}}\in\mathbb{R}^{T\times C}$, $W$ becomes $\bm{\hat{W}} \in\mathbb{R}^{K\times C}$ and $P$ becomes $\bm{\hat{P}} \in\mathbb{R}^{(t-1)\times C}$. Then the hidden states as intermediaries connect $\bm{\hat{X}}$, $\bm{\hat{W}}$ and $\bm{\hat{P}}$, through a multi-modal attention mechanism that is shown in Figure 2. Taking the acoustic information for example: given the previous LSTM hidden state $h_{t-1}$, we use a single fully-connected layer followed by a softmax function to generate the attention distributions $\alpha$ of acoustic features in the time dimension. Finally, the gated linear unit (GLU) is applied to the output of the attention module, to control how much information should flow into the next layer. Formula (3)-(5) are the definitions of the acoustic attention module $\bm{\Psi_{x}}$:
\begin{equation}
    \bm{A} = ReLU((\bm{\hat{X}}\bm{W_{i}^T}+b_{i}) \oplus  (h_{t-1}\bm{W_{s}^T}+b_{s}))   
\end{equation}
\begin{equation}
    \alpha = softmax(\bm{A}\bm{W_{n}}+b_{n})   \vspace{1ex}
\end{equation}
\begin{equation}
    o_{x} =  GLU([\bm{\hat{X}} \otimes \bm{\alpha},h_{t-1}])  \vspace{1ex}
\end{equation}
where $\bm{W_{s}}\in \mathbb{R}^{M\times{H}}$,  $\bm{W_{i}}\in \mathbb{R}^{M\times{C}}$, $\bm{W_{n}}\in \mathbb{R}^{M}$ are transformation matrixes that map acoustic features and hidden states to the same dimension. Here are $b_{s}\in\mathbb{R}^{M}$, $b_{i}\in\mathbb{R}^{M}$, and $b_{n}\in\mathbb{R}^{1}$. We denote $\oplus$ as the element-wise addition of a matrix and a vector, and $\otimes$ as the element-wise multiplication of a matrix and a vector. We choose the GLU operation to obtain the output $o_{x}\in\mathbb{R}^{C}$, which implements a simple gating mechanism over the output $\mathcal Y = [\mathcal A,\mathcal B]\in\mathbb{R}^{2d}$:
\begin{equation}
    GLU([\mathcal A, \mathcal B]) = \mathcal A \otimes \sigma(\mathcal B)
\end{equation}
where $\mathcal A\in\mathbb{R}^d, \mathcal B\in\mathbb{R}^d$ are the inputs to the non-linearity, and the output $GLU([\mathcal A,\mathcal B])\in\mathbb{R}^d$ is half the size of $\mathcal Y$ \cite{gehring2017convolutional}. 

As for the semantic information, the same structure of the attention module is applied to keywords and previously predicted words, and the outputs are $o_{w}\in\mathbb{R}^{C}$ and $o_{p}\in\mathbb{R}^{C}$ respectively. Note each part of semantic information shares an attention module. We add $o_{x}$, $o_{w}$, $o_{p}$ with $w_{t-1}$ which is a predicted word of the last time step to obtain the output $o_{t}$. Then, $o_{t}$ and $h_{t-1}$ are sent to calculate the hidden state $h_{t}$ which is used to predict word probability distribution $v_{t}$. Finally, the current word $w_t$ is chosen from $v_{t}$ with the highest probability and added to previously predicted words $\bm{P}$ for the next iteration of LSTM. Formula (7) is the operation of the multi-modal attention module described above:

\begin{equation}
    \begin{aligned}
    h_0 & = GAP(\bm{\hat{X}}) \\
    h_t &= LSTM(h_{t-1},Add(o_{x},o_{w},o_{p},\textbf{Emb}(w_{t-1}))) \\ 
    v_t &= Softmax(Linear(h_t))
    \end{aligned}
\end{equation}

\begin{figure*}[t]
  \includegraphics[width=\linewidth]{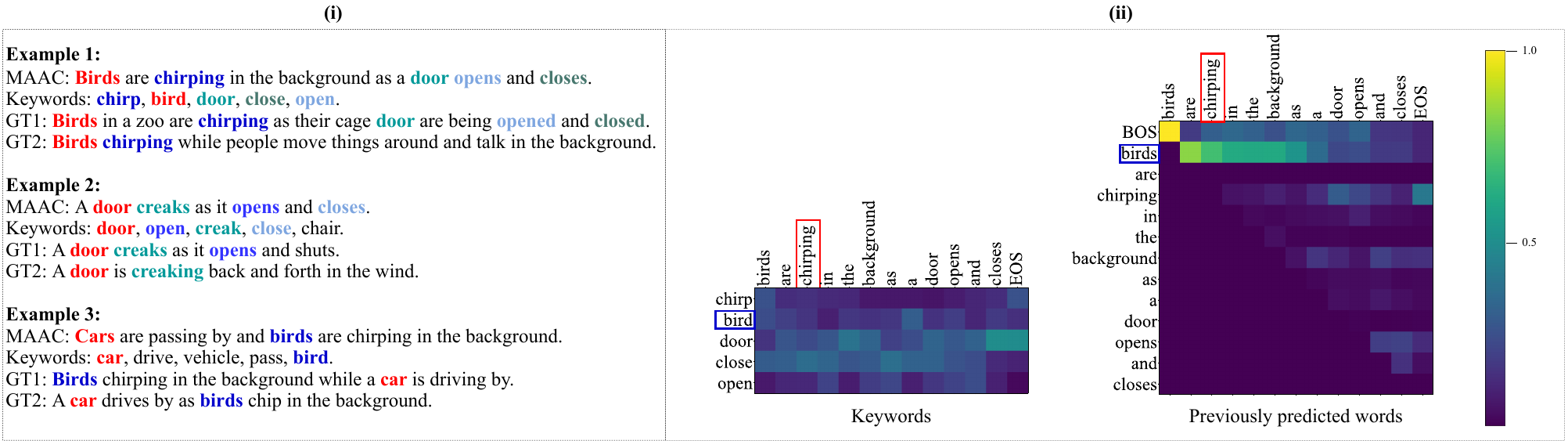}
  \caption{(i) It shows some examples of MAAC outputs and colored words indicate that keywords appear in both predicted and ground-truth sentences. (ii) The visualization for attention matrices of keywords and previously predicted words in the semantic attention module of example 1.}
  \label{fig:results2}
\end{figure*}
\noindent where $h_0$ represents the global information of acoustic features in the time dimension. $v_{t}\in\mathbb{R}^{|\Sigma|}$ is a probability vector, and $|\Sigma|$ is a predefined dictionary including all words.

\section{Experiment}
\label{sec:Experiment}

\subsection{Dataset and Experiment Setup}
\label{sec:Dataset and Experiment Setup}

\textbf{Clotho v2} We evaluate our proposed method on the Clotho v2 dataset \cite{drossos2020clotho}, which is published in DCASE 2020 and expanded in DCASE 2021. Nowadays it contains 5,929 audio clips labeled with 5 captions for each, including 3,839 training, 1,045 validation, and 1,045 testing audio clips. We convert all sentences to lower case and remove all punctuation marks, ending up with a vocabulary $|\Sigma|$ of 4368 words including special tokens "BOS", "EOS", and "PAD". For evaluation, we employ standard evaluation metrics: BLEU \cite{papineni2002bleu}, ROUGE-L \cite{lin2004rouge}, METEOR \cite{banerjee2005meteor}, CIDEr-D \cite{vedantam2015cider}, SPICE \cite{anderson2016spice} and SPIDEr that is the mean of CIDEr-D and SPICE. All metrics are computed with the audio captioning evaluation tool\footnote{\url{https://github.com/audio-captioning/caption-evaluation-tools}}.

\noindent\textbf{Implementation Details} We choose $N=300$ keywords for pre-training encoder, $K=5$ keywords and the dimension of fully-connected layers $C_{1}$, $C_{2}$ and $C$ are 512. The decoder LSTM has 512 hidden units, word embedding size is also set to 512. To mitigate overfitting, dropout regularization \cite{srivastava2014dropout} is used in the word embedding layer with a rate of 0.5, and the word classification layer with a rate of 0.25. 

The training strategy of the MAAC could be divided into two stages: encoder pre-training and the whole MAAC model training. In the phase of training the encoder, firstly the CNN backbone is frozen up, trained with the initial learning rate of $1\times10^{-3}$ for 80 epochs. Next, we finetune the whole keyword encoder with the learning rate of $5\times10^{-4}$ for 25 epochs. Then, it can be divided into two parts for training the whole MAAC: CE training and RL fine-tuning. CE training takes 30 epochs while the parameters of the pre-trained keyword encoder are frozen. Finally, the $30^{th}$ CE training model is used for reinforcement learning (RL) fine-tuning 55 epochs. In all training stages, we adopt an Adam optimizer with a mini-batch size of 32, and exponential decay to adjust the learning rate with a factor of 0.98 every epoch. The initial learning rates are set to $3\times10^{-4}$ and $5\times10^{-5}$ for two parts of training the whole MAAC. In the inference stage, we adopt beam search with a beam size of 4 that is implemented to achieve the best decoding performance.

In order to avoid over-fitting and increase data diversity, SpecAugment \cite{park2019specaugment}, SpecAugment++ \cite{wang2021specaugment++}, Mixup \cite{zhang2017mixup}, Label smoothing \cite{szegedy2016rethinking} and teacher forcing \cite{NIPS2015_e995f98d} are used in the training phase. For Mixup method, it is just used in the training of the keyword encoder. The label smoothing and teacher forcing are just used while training the whole MAAC. 
\subsection{Result Analysis}
\label{sec:Result Analysis}
We compare our proposed MAAC with the following current models: (1) Baseline \cite{drossos2020clotho} is proposed by K. Drossos \textit{et al.}, which employs a GRU-GRU encoder-decoder framework; (2) Temporal attention model (TAM) \cite{wang2020automated} uses the CNN encoder and the LSTM-based decoder with the temporal attention mechanism; (3) Transformer-based model (TM) \cite{wu2020audio} adopts a pre-training strategy to improve captioning performance; (4) UNIS's model \cite{xinhao2021_t6} uses PANNs to initialize the parameters of the encoder and is pre-trained on AudioCaps dataset \cite{kim2019audiocaps}; (5) SJTU's model \cite{xu2021_t6} utilizes AudioSet to pre-train its encoder in order to enhance the ability of the encoder to recognize audio concepts. Both (5) and (6) adopt RL training to obtain the final models. 

Table 1 lists the results of various single models on the Clotho dataset. Our MAAC achieves the highest score on all metrics in the CIDEr-D optimization stage. In addition, the CIDEr-D score of the proposed MAAC improves from 41.9 to 49.1 after further optimizing CIDEr-D. 

Through Figure 2 (i), we can find that the pre-trained keyword encoder can almost recognize the main concepts \textit{i.e.} keywords (e.g. \textit{bird} and \textit{chirp} in example 1) of a given audio clip, and the keywords may appear in different states in the ground-truth captions and the predicted sentences. Figure 2 (ii) further shows that keywords and previously predicted words are concerned to generate the current word. For instance, when the decoder is generating ``chirping'', it pays more attention to the ``birds" in the previously predicted words but pays less attention to ``birds" in the keywords. That is to say, previously predicted words and keywords are complementary to each other in the semantic attention module. 
\subsection{Ablative Analysis}
\label{sec:Ablative Analysis}
To quantify the impact of the proposed multi-modal attention module, we compare our MAAC against a set of other ablated models with different settings. The results of various models are shown in Table 2. We firstly design the ``base" model which does not use previously predicted words and keywords (\textit{i.e.} the semantic attention module). Then we add the information of previously predicted words or keywords to the "base" model. We find that it has little impact on the performance of the model by only introducing previously predicted words. It might be that previously predicted words would contain wrong words that destroys the input information of the decoder. In addition, the model which only uses the keywords in the semantic attention module could not converge. From section \ref{sec:Result Analysis}, we know that keywords contain the main concepts of an audio clip. When we only utilize them in the semantic attention module, they will cause the decoder to pay more attention to the part of the keywords and ignore the overall semantic relationship. Moreover, we examine the performance of using a shared (or not) semantic attention module on its performance and find that a sharing semantic attention module could further improve the CIDEr-D score.

\section{Conclusion}
\label{sec:conclusion}
In this paper, we propose a novel audio captioning model based on the multi-modal attention module which utilizes both acoustic and semantic information to generate captions. In addition, the performance of the MAAC achieves a new state-of-the-art under the two stages of training. The ablation experiments further demonstrate the effectiveness of the multi-modal attention module. In future work, we would concentrate on how to align the multi-modal information more effectively to improve the performance of the AAC.

\bibliographystyle{IEEEtran}
\bibliography{refs}

%
%
%
%
%
%
%
%
%

\end{sloppy}
\end{document}